\documentclass[conference]{IEEEtran}
\IEEEoverridecommandlockouts
\usepackage{cite}
\usepackage{amsmath,amssymb,amsfonts}
\usepackage{graphicx}
\usepackage{textcomp}
\usepackage{xcolor}
\def\BibTeX{{\rm B\kern
-.05em{\sc i\kern-.025em b}\kern-.08em
    T\kern-.1667em\lower.7ex\hbox{E}\kern-.125emX}}
\usepackage{graphicx} 
\usepackage{amsmath,amsfonts}
\usepackage{algorithm}
\usepackage{algpseudocode}
\usepackage{array}
\usepackage[caption=false,font=normalsize,labelfont=sf,textfont=sf]{subfig}
\usepackage{textcomp}
\usepackage{stfloats}
\usepackage{url}
\usepackage{verbatim}
\usepackage{graphicx}
\usepackage{cite}
\usepackage{dsfont}
\usepackage{xcolor}
\usepackage{subfig}
\usepackage{acronym} \input{Acronym}
\usepackage{pgfplots}
  \pgfplotsset{compat=newest}
  \usetikzlibrary{plotmarks}
  \usetikzlibrary{arrows.meta}
  \usepgfplotslibrary{patchplots}
  \usepackage{grffile}
  \usepackage{amsmath}

  \pgfplotsset{plot coordinates/math parser=false}
  \newlength\figureheight
  \newlength\figurewidth
\begin{document}
\makeatletter
\def\ps@IEEEtitlepagestyle{%
  \def\@oddfoot{\hfil \footnotesize 
  \parbox{6in}{\centering \textcopyright\ 2026 IEEE. Personal use of this 
  material is permitted. Permission from IEEE must be obtained for all other 
  uses, in any current or future media, including reprinting/republishing 
  this material for advertising or promotional purposes, creating new 
  collective works, for resale or redistribution to servers or lists, or 
  reuse of any copyrighted component of this work in other works.}
  \hfil}%
  \def\@evenfoot{}}
\makeatother
\title{Design and Practical Validation of a Novel Modulation Scheme for RIS Detection and Identification

}
\author{Aymen~Khaleel, ~\IEEEmembership{Member,~IEEE}, Adam~Umra, ~\IEEEmembership{Graduate Student Member,~IEEE}, and Aydin~Sezgin, \\~\IEEEmembership{Senior Member,~IEEE}
\vspace{-0.3cm}

\thanks{
This work was supported in part by the German Federal Ministry of Research, Technology and Space (BMFTR) in the course of the 6GEM+ Transfer Hub under grant 16KIS2411 and in part by the German Research Foundation (“Deutsche Forschungsgemeinschaft”) (DFG) under Project–ID 287022738 TRR 196 for Project S03.

Aymen Khaleel, Adam Umra,  and Aydin Sezgin are with the Faculty of Electrical Engineering and Information
Technology, Ruhr-University Bochum, 44801 Bochum, Germany (e-mail: \{aymen.khaleel, adam.umra, and
aydin.sezgin\}@rub.de).

}
}
\maketitle
\begin{abstract}
The reconfigurable intelligent surfaces detection and identification (RISs-ID) is a critical process that enables a base station (BS) to adaptively assign the appropriate RIS to a given \ac{UE}. This work proposes a novel modulation scheme to enhance the reliability of RIS-ID by reducing the miss-detection ($P_\text{m}$) and false-alarm ($P_\text{f}$) probabilities. Specifically, we leverage the RIS's passive beamforming gain to enable over-the-air modulation of the RIS ID, combined with passive beam sweeping to extend detection coverage in angular space. Prototype experiments demonstrate the effectiveness of the proposed scheme in reducing $P_\text{m}$ and $P_\text{f}$ to $10^{-3}$ across different operating points.


\end{abstract}


\section{Introduction}
{R}{econfigurable} intelligent surfaces (RISs) are constantly receiving attention from the wireless research society as an enabling technology for sixth-generation (6G) of wireless communication systems \cite{RISfor6G}. An RIS comprises a large array of electronically adjustable elements to control the amplitude, phase, and polarization of the signals hitting their surfaces, enabling the reshaping of the propagation environment for extended coverage, enhanced data rates, and increased energy efficiency\cite{RIS_Advantages}. RISs are envisioned to be deployed in large numbers at the base station (BS) or user equipment (UE) side, as the two most practical scenarios to assist the BS-UE wireless communications \cite{ris-deploy}.



Recently, the authors in \cite{RIS_ID_Theoretical} investigated the RIS detection and identification (RIS-ID) problem and proposed a complete mathematical solution. Specifically, the authors considered the problem of a BS  attempts to detect and identify the UE-RIS-BS link status (blocked or not). This is an initial procedure before the BS can optimize the RIS phase configuration to serve that UE. One of the essential steps in the RIS-ID is the over-air-modulation step that needs to be done at the RIS side. In this context, the authors in \cite{RIS_ID_Theoretical} proposed to use \ac{BPSK} modulation by changing all of the RIS elements' phases between $0$ and $\pi$ according to the binary code sequence (BCS) associated with the RIS unique ID. However, this requires synchronization between the UE-BS in advance; otherwise, the phase and frequency offsets distort the received BPSK symbols. To relax the synchronization condition, the authors in \cite{ris-id-pract1} proposed a novel \ac{ASK} modulation method by relying on the phase-dependent amplitude variations associated with the RIS phase adjustment \cite{abeywickrama2020intelligent}. Furthermore, the authors in \cite{ris-id-pract1} validated the full RIS-ID concept proposed in \cite{RIS_ID_Theoretical} and their proposed modulation scheme through a lab experiment, showing its potential in a real-world environment. In \cite{RIS-ID-beam}, authors extended the RIS-ID concept into a more comprehensive protocol by considering a novel procedure for RIS-ID and beamforming. Specifically, the authors proposed a novel partitioning scheme to allocate different elements of an RIS for identification and passive beamforming simultaneously. 

However, the proposed modulation scheme in \cite{ris-id-pract1} has two limitations, as follows. First, due to the limited range in the amplitude variations associated with the phase change \cite{abeywickrama2020intelligent}, the created \ac{ASK} symbols have a small amplitude separation between the two ASK levels, creating higher decoding errors at the receiver side. Second, since all of the elements in \cite{ris-id-pract1} have a common phase shift at a time (uniform adjustment), the angle of reflection is always equal to the angle of incidence (specular reflection). This creates coverage holes on the UE side at the locations outside the ones covered by the angle of reflection. Overall, both aforementioned limitations result in higher miss-detection and false-alarm probabilities, reducing the reliability of the RIS-ID at the BS side.


Against the above background, we propose a novel passive beamforming-based modulation scheme to generate \ac{ASK} modulation symbols with a higher separation distance in the signal space and with wider reflection coverage. Specifically, leveraging the high passive beamforming gain the RIS can provide, we propose to create two received signal levels by either constructively or destructively aligning the signal at the receiver side. Thus, our proposed scheme can achieve a much larger symbol separation distance, enabling more reliable decoding at the receiver side. Furthermore, we combine passive beam sweeping with our proposed modulation scheme to enable the RIS to modulate and reflect the incoming UE signal from a wider range of incidence angles, unlike the case in \cite{ris-id-pract1}.

The rest of the paper is organized as follows: Section II first revisits the RIS-ID concept, followed by the introduction of the system model and the proposed modulation scheme. Section III provides the simulation results, followed by the experimental setup and obtained measurements. Finally, Section IV concludes the paper.
\section{System Model}
In what follows, we first briefly revisit the RIS-ID concept and its state-of-the-art solution, then introduce our novel modulation scheme and the corresponding system model. 
\begin{figure}[t!]
  \centering
   \includegraphics[width=0.9\columnwidth]{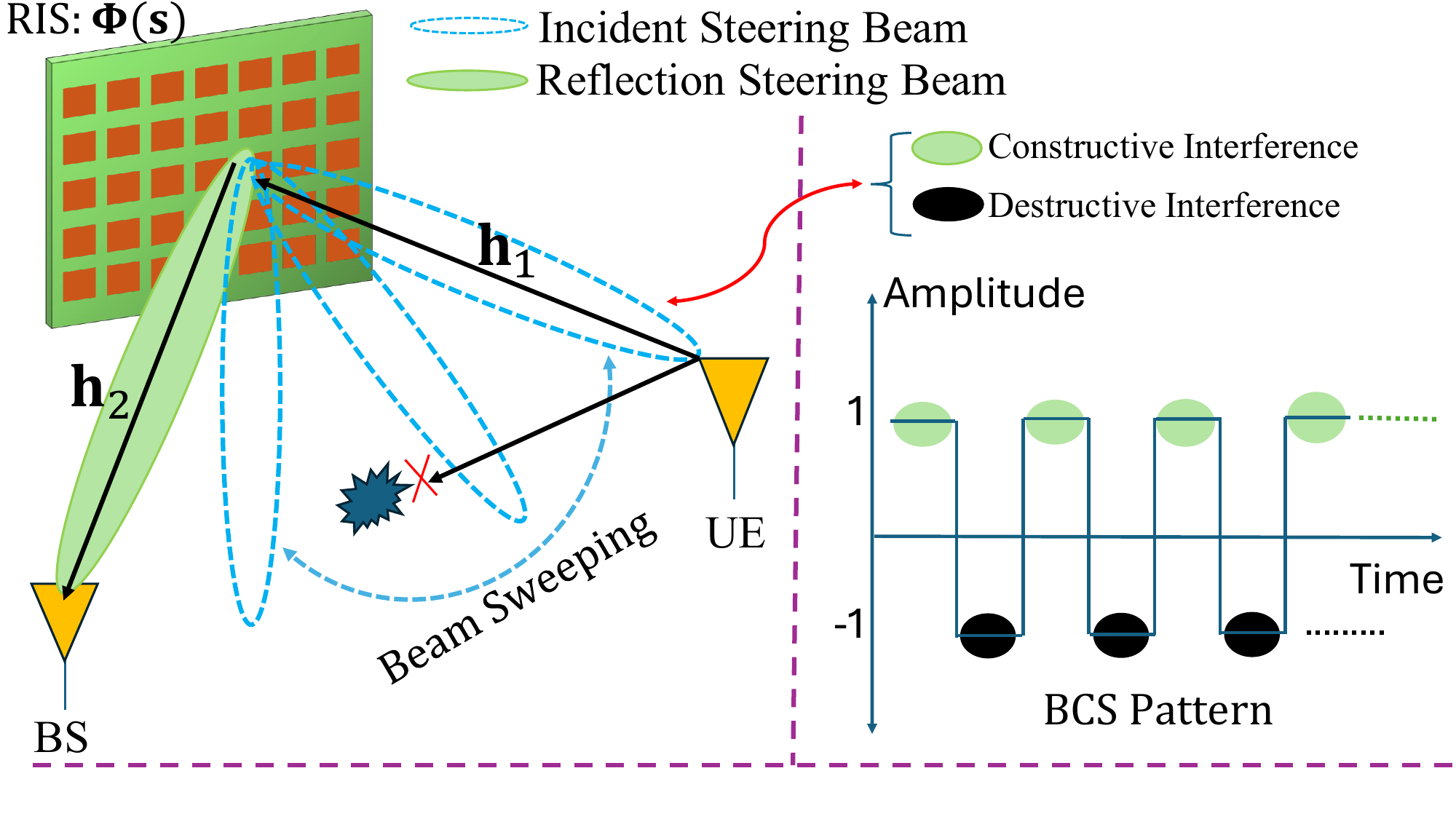}
   \caption{The proposed modulation scheme for the RIS-ID process. Here, the RIS adjusts its phase shift, $\mathbf{\Phi(\mathbf{s})}$, as a function of a BCS $\mathbf{s}$, associated with its unique ID.}\label{fig:sys-block}
   \vspace{-0.3cm}
\end{figure}

\subsection{ Revisiting the RIS Detection and Identification Process}

As proposed in \cite{RIS_ID_Theoretical}, the RIS-ID process involves the UE continuously transmitting an unmodulated carrier signal in the uplink to enable the \ac{BS} to determine whether the UE-RIS-BS link is blocked or not. Note that this process is performed prior to channel estimation \cite{rev1,rev2}, as the latter requires the UE-RIS-BS link availability information. Independent from the UE transmission and in a periodic manner, each RIS in the environment adjusts its phase shifts to over-the-air modulate the impinging signal with a BCS associated with its unique ID. Using a look-up table that contains RISs IDs and their corresponding BCSs, the BS correlates the received signal with all of the BCSs in the look-up table, and if the correlation amplitude exceeds a predefined threshold, the corresponding RIS is declared reachable: the UE-RIS-BS link is not blocked. Accordingly, the RIS-ID process is an initial step before the BS can optimize the phase configuration of each RIS deployed in the environment to serve a specific UE. 
\vspace{-0.1 cm}
\subsection{ Proposed Modulation Scheme}
Consider a single-antenna \ac{UE} sending unmodulated pilot signals in the uplink to enable a single-antenna \ac{BS} to determine whether the UE-RIS-BS link is blocked or not, as shown in Fig. \ref{fig:sys-block}. If the link is not blocked, then the signal that is over-the-air modulated and reflected by the RIS is received at the BS side as \cite{RIS_ID_Theoretical}
\vspace{-0.1 cm}
\begin{align}
y_m &=\sqrt{P_t}\left[\mathbf{h}_{2}^T \mathbf{\Phi}(s_m^{(i)}) \mathbf{h}_{1}\right]x+n_m,\\
&=\sqrt{P_t}h_{e}(s_m^{(i)})x+n_m,
\label{eq:y} 
\end{align}
where, $m$ is the BCS symbol index\footnote{The index $m$ is omitted from the channel vectors assuming the channel coherence time is larger than the transmission time required to send the full BCS. Otherwise, the correlation process can be segmented to realize this condition \cite{synch-book}.}, $x=1$ corresponds to the baseband sample of an unmodulated carrier signal sent by the UE with a transmit power $P_t$. The effective channel observed at the BS side is $h_{e}(s_m^{(i)})=\mathbf{h}_{2}^T \mathbf{\Phi}(s_m^{(i)}) \mathbf{h}_{1}$, where $\mathbf{h}_{1}, \mathbf{h}_{2}\in\mathbb{C}^{N\times 1 }$ are UE-RIS and RIS-BS channel vectors, respectively, with $N$ denoting the RIS size; $n_m\sim\mathcal{N}_\mathbb{C}(0,\sigma_n^2)$ is the additive white Gaussian noise (AWGN) sample with  variance $\sigma_n^2$ and $\mathbf{\Phi}(s_m^{(i)})\in\mathbb{C}^{N\times N}$ denotes the RIS phase shift matrix as a function of a BCS symbol $s_m^{(i)}$ to be modulated over the impinging signal, see Fig. \ref{fig:sys-block}. Here, $s_m^{(i)}=(-1)^i, i\in\{1,2\}$, and the BCS $\mathbf{s}=[s_1^{(i)},\dots,s_m^{(i)},\dots, s_M^{(i)}]$ is associated with the RIS unique ID, where different RISs have different BCSs. Specifically, at each time slot $m$ and according to the current symbol of the BCS that needs to be reflected, the RIS applies one of the following two possible phase shift profiles:
\vspace{-0.2 cm}
\begin{align}
\mathbf{\Phi}(s_m^{(i)})=
\begin{cases} 
      \mathbf{\Phi}_d, & i=1, \\
      \mathbf{\Phi}_c, & i=2,
   \end{cases}\label{eq:mod}
   \quad \forall m.
\end{align}

Here, $\mathbf{\Phi}_d$ and $\mathbf{\Phi}_c$ are designed to ASK-modulate $x$ by creating two distinct signal amplitude levels associated with the two possible BCS symbols $s_m^{(1)}=-1$ and $s_m^{(2)}=1$. In this way, the BS can map the measured high and low signal amplitude levels to their corresponding BCS symbols $1$ and $-1$, respectively. Furthermore, to reduce the ambiguity on the BS side to distinguish between the two symbols, $\mathbf{\Phi}_c$ and $\mathbf{\Phi}_d$ are designed to achieve the maximum possible separation distance between the two signal amplitude levels. Specifically, $\mathbf{\Phi}_c$ and $\mathbf{\Phi}_d$ are designed to destructively and constructively align the reflected signals at the BS side, creating the required high and low signal amplitude levels, respectively, as follows.

Considering an RIS with only two possible phase shifts\footnote{This is due to the RIS hardware used in this experiment. Nevertheless, the same concept can be applied to any continuous or discrete phase shift RIS.}, we define
\vspace{-0.1 cm}
\begin{subequations}\label{prob1:main}
\begin{align}
    \mathbf{\Phi}_c=\;&\underset{\mathbf{\Phi}}{\arg\max}\quad|\mathbf{h}_{2}^T \mathbf{\Phi} \mathbf{h}_{1}|\label{prob2:obj}\\
   & \quad\text{s.t.} \quad[\mathbf{\Phi}]_{k,k}\in\{-1,1\}, \forall k\in\{1,\dots, N\},\label{prob1:const1}
    \end{align}
\end{subequations}\\
and
\begin{subequations}\label{prob2:main}
\begin{align}
    \mathbf{\Phi}_d=\;&\underset{\mathbf{\Phi}}{\arg\min}\quad|\mathbf{h}_{2}^T \mathbf{\Phi} \mathbf{h}_{1}|\\
   &\quad \text{s.t.} \quad[\mathbf{\Phi}]_{k,k}\in\{-1,1\}, \forall k\in\{1,\dots, N\},\label{prob2:const}
    \end{align}
    \end{subequations}
    where, $[ \mathbf{\mathbf{\Phi}}]_{k,k}=e^{j\theta_{k}}$  denotes the $k$-th diagonal element of the diagonal matrix  $\mathbf{\Phi}\in\mathbb{C}^{N\times N}$, with $\theta_{k}\in\{0, \pi\}$. Due to the non-convexity of the problems \eqref{prob1:main} and \eqref{prob2:main}, we consider the following sub-optimal, yet practical, solutions. For \eqref{prob1:main}, we relax the discrete phase constraint to obtain the  optimal continuous phase shift solution as $\tilde{\theta}_k=-\arg([\mathbf{h}_1]_k[\mathbf{h}_2]_k)$. 
    Next, the discrete phase shift solution can be obtained as 
\begin{equation} \theta_
k^*=\underset{\theta_k\in\{0,\pi\}}{\arg\min}|\tilde{\theta}_k-\theta_k|.\label{eq:phs}
\end{equation} 
To solve \eqref{prob2:main}, element-wise alternating optimization \cite{disc-phs} can be used as follows. Starting from an initial state, $\theta_k=0, \forall k$, each RIS element phase $\theta_k$ is sequentially optimized by evaluating candidate phase values ($0$ or $\pi$) while keeping the other elements fixed. The phase profile that minimizes $|\mathbf{h}_2^T\mathbf{\Phi}\mathbf{h}_1|$ is selected, and the procedure is repeated for all elements. 

At the BS side, to obtain the ASK-modulated received symbols centered around zero, similar to $s_m^{(i)}$, the envelope of the collected received samples, $\mathbf{b}[l]=|y_l|$, for $l\in\{1, \dots, L\}$, where $L$ is the number of collected samples, is mean-centered as
    \begin{equation}
\Tilde{\mathbf{b}} = \mathbf{b} -\bar{\mathbf{b}},  
    \end{equation}
where $\bar{\mathbf{b}}$ denotes the sample mean of $\mathbf{b}$. Next, the BS correlates $\Tilde{\mathbf{b}}$ with the BCSs
in a look-up table, as explained in Section I-A.
Without loss of generality, assuming a single deployed RIS, we obtain the correlation amplitude as
    \begin{equation}
        \mathbf{a}[n] = \frac{1}{M}\sum_{m=0}^{M-1} \mathbf{s}[m]\Tilde{\mathbf{b}}[m+n],\;-(L-1)\leq n\leq M-1,\label{eq:corr}
    \end{equation}
    where, $\mathbf{a}[n]$ denotes the $n$-th element of the vector $\mathbf{a}$, and $n$ is the correlation index. For the purpose of RIS detection, we search for the maximum peak, which is given as
    \begin{equation}
        a^*=\underset{n}{\max}\;\left|\mathbf{a}[n]\right|.\label{eq:corr2}
    \end{equation}
  Next, $a^*$ is compared to the detection threshold to provide a decision: the UE-RIS-BS link is blocked or not, as elaborated in \cite{RIS_ID_Theoretical}. In this context, our proposed novel modulation scheme, expressed in \eqref{eq:mod}, aims to maximize $a^*$ through boosting the detection of the ASK symbols, and thus, reducing the miss-detection and false-alarm probabilities \cite{RIS_ID_Theoretical}, \cite{ris-id-pract1}.

\subsection{Passive Beam Sweeping}
One of the main issues with \cite{ris-id-pract1} is the signal coverage limited by the incident angle, due to the lack of passive beamforming, as explained in Section I. In this work, a passive beam sweeping method is used to increase the RIS-ID coverage, as follows. 


The angular space within the field of view (FoV) of the RIS, associated with the potential incident directions from the UE, is sampled into a finite set of steering vectors using a Discrete Fourier Transform (DFT) codebook of size $V$.
Next, the RIS applies Algorithm 1, as follows.  While $\mathbf{h}_2$, and thus the reflection beam, is fixed due to the BS fixed location, $\mathbf{h}_1$ is recalculated based on the $v$-th steering vector.
Accordingly, the corresponding constructive and destructive beams, $\mathbf{\Phi}_c$ and $\mathbf{\Phi}_d$, are calculated 
 and applied sequentially over time according to the RIS BCS, see Fig. \ref{fig:sys-block}. Thus, by applying Algorithm 1, the RIS reflects the full BCS associated with its unique ID from all of $V$ incident angular directions in its FoV to the BS, extending the coverage range of the RIS-ID process over a significantly wider range of UE potential angular directions. Note that $\mathbf{h}_1$ can be obtained from the steering vector in the adopted DFT codebook, while $\mathbf{h}_2$ can be readily computed using the geometry of 
line-of-sight (LoS) channels, as both the RIS and BS have known and fixed locations.

Considering the computational complexity, for solving  \eqref{prob2:main} and \eqref{eq:phs}, a linear computational cost of $\mathcal{O}(N)$ is required. Note that the RIS, irrespective of the UE location, applies Algorithm 1 periodically to scan its FoV. Therefore, $\mathbf{\Phi}(s_m^{(i)})$ can be obtained once for a given DFT codebook. In this way, the execution time associated with Algorithm 1 scales with the DFT codebook size and the BCS length, $\mathcal{O}(VM)$. This means that, for a given RIS phase switching frequency, 
$M$ must be selected carefully to ensure that the channel 
coherence time associated with UE mobility is always greater 
than the time required to perform a full DFT codebook scan. 
Otherwise, the UE may fail to receive all $M$ symbols of 
the BCS, which can effectively degrade the correlation 
amplitude.
\begin{algorithm}[t!]
\caption{Passive Beam Sweeping}
\begin{algorithmic}[1]
\State \textbf{Input:}  $V$, $\mathbf{s}$, $\mathbf{h}_2$.
\For{$v = 1$ to $V$}
\State \hspace{-0.6cm} Calculate $\mathbf{h}_1$ based on the $v$-th steering direction.
\State  \hspace{-0.5cm}Solve \eqref{prob1:main} and \eqref{prob2:main}, based on the new $\mathbf{h}_1$, to obtain $\mathbf{\Phi}_c$ and $\mathbf{\Phi}_d$, respectively.
      \For{$m = 1$ to $M$}
        \State  Set the RIS phases as $\mathbf{\Phi}(s_m^{(i)})\in\{\mathbf{\Phi}_c,\mathbf{\Phi}_d\}$, as given in \eqref{eq:mod}.
     \EndFor
\EndFor
\end{algorithmic}
\end{algorithm}
\section{Simulation and Experimental Results}
This section first provides the simulation results and discusses them. Next, the experiment conditions and parameters are introduced, followed by the reported measurements and their discussion.
\subsection{Simulation Results}
To assess the performance of the proposed ASK modulation scheme, \eqref{eq:mod}, the performance metric used is the difference in amplitude between the two effective channels observed at the BS side, which are associated with the two symbols of the BCS: $d=||h_e(s_m^{(2)})|-|h_e(s_m^{(1)})||$. The higher the value of $d$, the more reliable the decoding of the two symbols at the BS side, and thus, the higher the correlation amplitude $a^*$. 

\begin{figure}[t!]
  \centering
    \hspace*{-2em}
   \includegraphics[width=70mm,height=50mm]{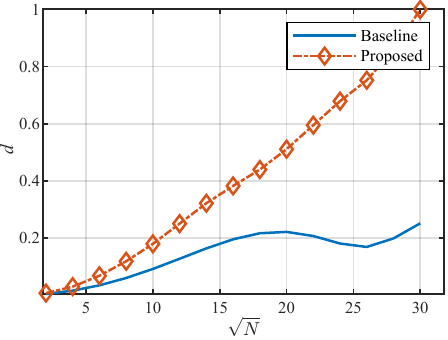}
   \caption{The amplitude difference $d$ (normalized) versus the RIS size, under full LoS channel conditions.}\label{fig:amp_diff}
\end{figure}

Fig. \ref{fig:amp_diff}, shows that the proposed scheme surpasses the baseline \cite{ris-id-pract1} with a substantial performance margin that increases with the RIS size. This is due to the passive beamforming gain, which is a quadratic function in $N$ \cite{ris-quad}. In contrast, the baseline scheme achieves no passive beamforming as the RIS elements are adjusted to the same common phase shift at a time, creating non-coherent destructive and constructive alignment of the reflected signals on the BS side. As a result, the baseline exhibits the non-monotonic behavior shown in the figure.

\begin{figure}[t!]
    \centering
    \includegraphics[width=\columnwidth]{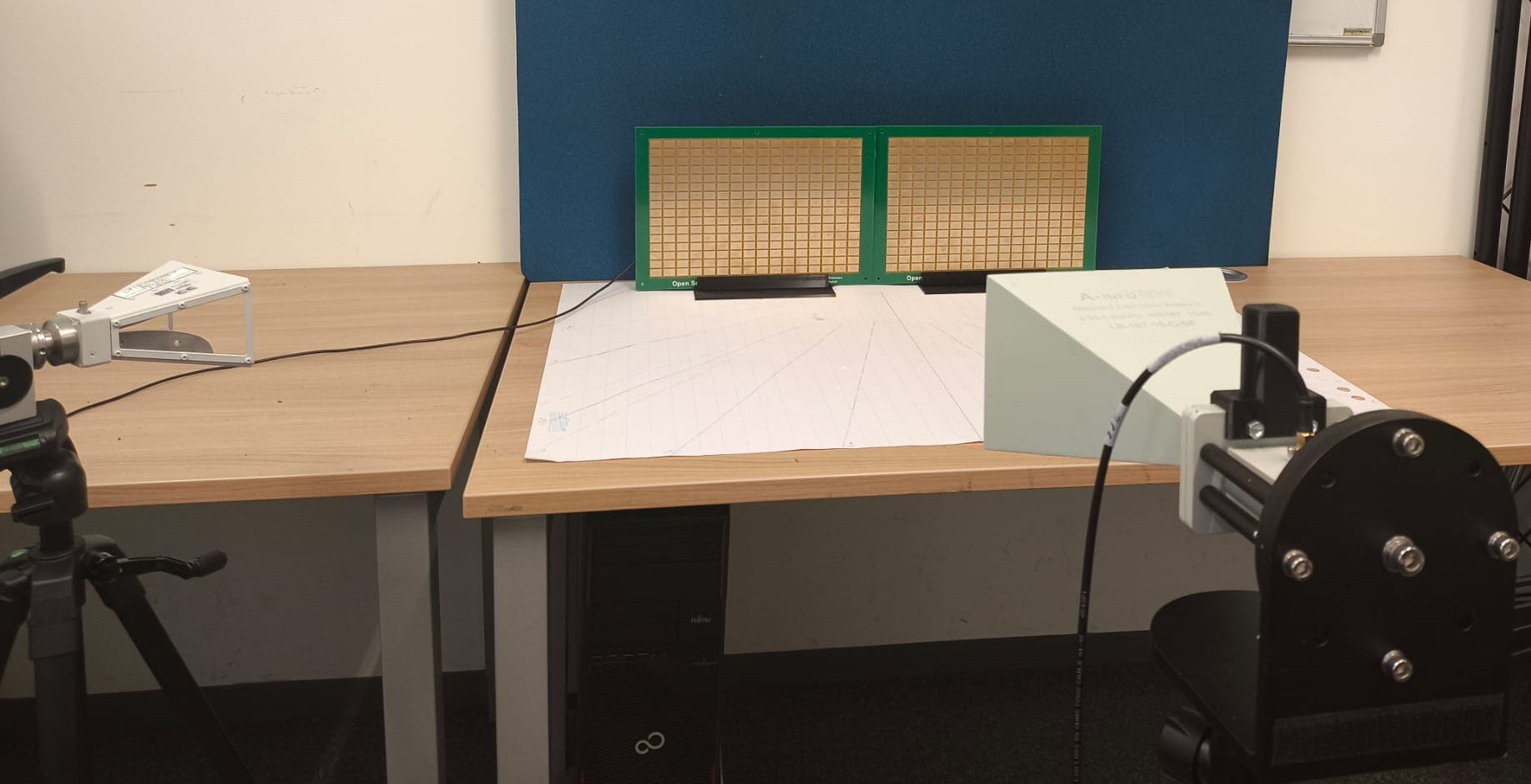}
    \caption{The experiment setup shows two RISs with two horn antennas, one for the UE and the other for the BS. }\label{fig:setup}
\end{figure}
\label{Sec:ExpSetup}
\subsection{Experiment Setup}
In Fig. \ref{fig:setup}, the experimental setup is shown, where ADALM-PLUTO software-defined radio (SDR) modules \cite{pluto-sdr} are used with horn antennas for both the transmitter (UE) and the receiver (BS).
Note that, unless otherwise specified, the two RISs are used and optimized as a single RIS. The two SDRs and RISs are all connected to a host computer and controlled using MATLAB and Python. 
The experimental configuration employs two RISs, each comprising $N_1 = N_2 = 256$ elements with 1-bit phase control \cite{ris-module}, with the UE transmission gain and BS receiving gain set to $0\,\mathrm{dB}$ and $30\,\mathrm{dB}$, respectively. The system operates at $5.53\,\mathrm{GHz}$ in alignment with the optimum operating frequency of the used RIS \cite{ris-module}. A sampling frequency of $2.6\,\mathrm{MHz}$ is used to provide adequate time resolution for real-time signal processing while maintaining computational efficiency on the host computer. Horn antennas with $15\,\mathrm{dBi}$ gain are used at the BS and UE sides. Considering the indoor office environment, the BS--RIS and UE--RIS distances are set to $2\,\mathrm{m}$ and $0.6\,\mathrm{m}$, respectively. The BCS length is set to $M = 16$ to balance RIS switching speed constraints with sufficient correlation peak magnitude to exceed the noise floor by a noticeable margin, as shown in Fig. \ref{fig:ave1}.

\subsection{Experiment Results}
In this subsection, we report the experimental measurements for our proposed and baseline schemes. Specifically, we compare both schemes regarding the ASK modulation symbols they generate: the difference between the two amplitude levels. Furthermore, we implement the RIS-ID process at the BS side to measure the correlation amplitude associated with both schemes. Finally, to evaluate the resulting RIS-ID performance, two metrics are considered: the false alarm probability and miss-detection probability, $
P_\text{f} = P(a^* > \bar{r} \mid \text{RIS is not present})$,
 $P_\text{m} = P(a^* < \bar{r} \mid \text{RIS is present})$, respectively, where $\bar{r}=r/\sigma_n$ and $r=a^*$ is obtained using (\ref{eq:corr2}) when there is no transmission from the UE side [7]. In practice, the detection threshold and the BCS length can be obtained for a given $P_\text{f}$, and $P_\text{m}$ can then be computed accordingly \cite[Eqs.~(14) and~(22)]{RIS_ID_Theoretical}.

\begin{figure}[t!]
    \centering
   \includegraphics[width=\columnwidth]{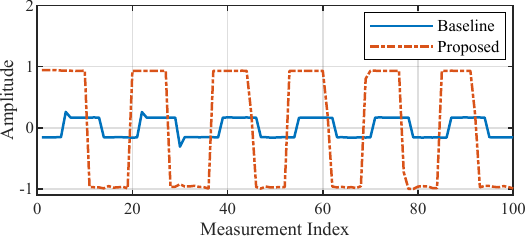}
    \caption{Amplitude levels (normalized) associated with the two BCS symbols $1$  and $-1$.}\label{fig:amp1}
\end{figure}

In Fig. \ref{fig:amp1}, we show the received signal amplitude levels for the proposed and baseline schemes, where it can be clearly seen that the proposed scheme achieves a remarkable performance gain in maximizing the amplitude difference between the two ASK symbols ($1$ and $-1$). 

\begin{figure}[t!]
    \centering
   \includegraphics[width=\columnwidth]{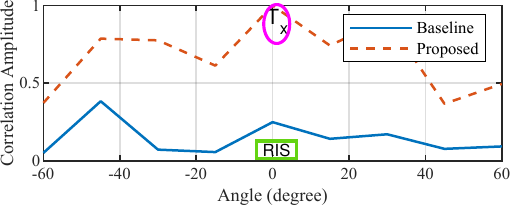}
    \caption{Correlation amplitudes $a^*$ (normalized) for the proposed and baseline scheme over a range of angles with respect to the RIS center point. Here, the UE is in a fixed position while the BS changes its position on the angle axis.}\label{fig:ave1}
     \vspace{-0.5 cm}
\end{figure}

In Fig. \ref{fig:ave1}, the correlation amplitudes measured at the BS side are shown, where the proposed scheme significantly outperforms the baseline across the measured angle range. This can be explained by the high passive beamforming gain achieved due to the phase adjustment using (\ref{eq:phs}). In contrast, the baseline scheme exhibits specular reflection — reflecting the signal toward the angle equal to the incidence angle — due to the uniform phase adjustment, with no passive beamforming gain. Here, the average amplitude (normalized) achieved by the proposed and baseline schemes is $1$ and $0.212$, respectively. This shows that the proposed scheme can provide much higher coverage for detecting the RIS, over a significantly wider range of UE and BS angular directions within the RIS FoV.
\begin{figure}[t!]
    \centering
   \includegraphics[width=88mm]{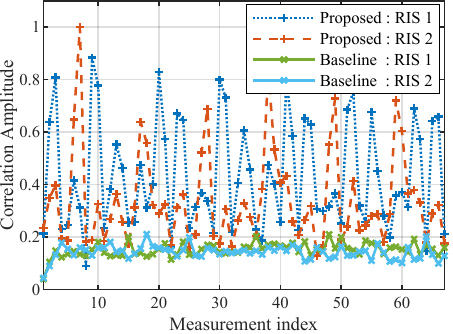}
    \caption{Correlation amplitudes $a^*$ (normalized) for the proposed and baseline schemes when Algorithm 1 is applied, considering two different RISs.}\label{fig:beamswep}
\end{figure}

In Fig. \ref{fig:beamswep}, we consider the case of two different RISs (each with $N=256$), where a different BCS is assigned to each RIS module. Furthermore, each RIS module employs Algorithm 1 to enable its detection at the BS side by sweeping its FoV with the passive constructive/destructive beams according to its unique BCS. It can be seen that the proposed scheme outperforms the baseline by a significant margin with each RIS module. 

\begin{figure}[t!]
    \centering
   \includegraphics[width=\columnwidth]{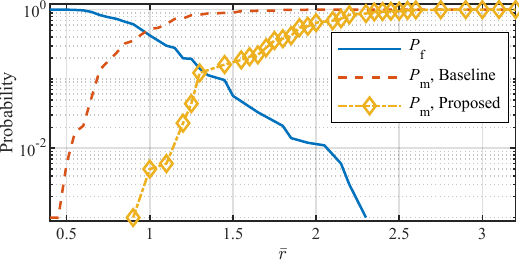}
    \caption{RIS Detection performance under full LoS conditions for both the UE–RIS and RIS–BS channels.}\label{fig:prob_NoBlock}
     \vspace{-0.5 cm}
\end{figure}

\begin{figure}[t!]
    \centering
   \includegraphics[width=\columnwidth]{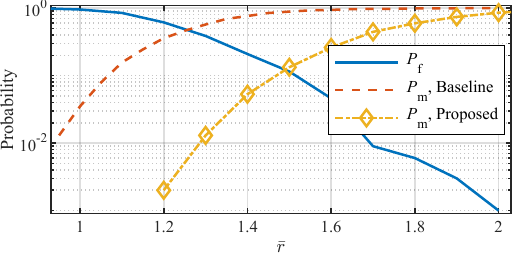}
    \caption{RIS Detection performance under blocked LoS conditions
for both the UE–RIS and RIS–BS channels. In this setup, the horn antennas are replaced with dipole antennas to enable omnidirectional reception of the reflected signals.}\label{fig:prob_block}
     \vspace{-0.5 cm}
\end{figure}

In Fig. \ref{fig:prob_NoBlock}, we show $P_\text{m}$ and $P_\text{f}$ calculated based on $1000$ different channel realizations, for statistical reliability. Here, as $\bar{r}$ increases $P_\text{f}$ decreases while $P_\text{m}$ increases, which aligns with the theoretical results obtained in \cite{RIS_ID_Theoretical}. Furthermore, in terms of $P_\text{m}$, the proposed scheme shows a clear performance gain over the baseline across the full  range of $\bar{r}$, which aligns with the results in Figs. \ref{fig:ave1}-\ref{fig:beamswep}. Note that, due to the strong LoS link, a shorter BCS ($M = 7$) is considered to induce miss-detection events. A similar performance is observed in Fig. \ref{fig:prob_block}, where the RIS-ID scheme enables RIS detection even in the absence of a LoS link, i.e., based solely on environmental reflections, which is not reported in \cite{ris-id-pract1}. 
\section{Conclusion}
In this work, we proposed a novel modulation scheme for the RIS-ID process that leverages the RIS's high constructive/destructive passive beamforming, combined with passive beam sweeping, to reduce the miss-detection and false-alarm probabilities and widen the coverage area. The proposed modulation scheme has been validated through simulations and practical experiments, demonstrating a clear performance gain over the state-of-the-art baseline. The obtained results show a high potential for the proposed scheme to extend the detection coverage of RISs and make it more reliable.

\bibliographystyle{IEEEtran}
\bibliography{refs}

\end{document}